\documentclass[reprint,prx,showpacs,twocolumn,superscriptaddress,aps,nofootinbib]{revtex4-2}

\usepackage[utf8]{inputenc}
\usepackage[american,british]{babel}
\usepackage[T1]{fontenc}
\usepackage[pdftex]{graphicx}  
\usepackage{graphicx}
\usepackage{xcolor}
\usepackage{mathtools}
\usepackage{bbm}
\usepackage{subfigure}
\usepackage{dcolumn}
\usepackage{braket}
\usepackage{bm}
\usepackage{amsmath,amsthm,amssymb}
\usepackage{color}
\usepackage{verbatim}
\usepackage{comment}

\definecolor{Red}{HTML}{E53E30}  
\definecolor{Green}{HTML}{00AD69} 
\definecolor{Blue}{HTML}{2171b5}
\definecolor{Purple}{HTML}{652F6C}

\usepackage[colorlinks,citecolor=blue,linkcolor=blue,urlcolor=blue,hyperindex]{hyperref}
\usepackage[dvipsnames]{xcolor}

\usepackage[normalem]{ulem}

\usepackage[nameinlink,noabbrev]{cleveref}
\makeatletter
\def\@seccntformat#1{\@ifundefined{#1@cntformat}%
   {\csname the#1\endcsname\quad}
   {\csname #1@cntformat\endcsname}
}
\makeatother
\DeclareMathAlphabet\mathbfcal{OMS}{cmsy}{b}{n}

\usepackage{dsfont}
\usepackage{physics}

\begin{document}

\title{Network theory classification of quantum matter based on wave function snapshots}

\author{Riccardo Andreoni}
\altaffiliation{These authors contributed equally}
\email{randreon@sissa.it}                        
\affiliation{International School for Advanced Studies (SISSA), via Bonomea 265, 34136 Trieste, Italy}
\affiliation{The Abdus Salam International Centre for Theoretical Physics (ICTP), strada Costiera 11, 34151 Trieste, Italy}

\author{Vittorio Vitale}
\altaffiliation{These authors contributed equally} 
\email{randreon@sissa.it} 
\affiliation{Pasqal, 24 rue Emile Baudot, 91120 Palaiseau, France}

\author{Cristiano Muzzi}
\affiliation{International School for Advanced Studies (SISSA), via Bonomea 265, 34136 Trieste, Italy}
\affiliation{INFN, Sezione di Trieste, via Valerio 2, 34127 Trieste, Italy}
\affiliation{International Solvay Institutes, 1050 Brussels, Belgium}
\affiliation{Center for Nonlinear Phenomena and Complex Systems, Universit\'e Libre de Bruxelles, CP 231, Campus Plaine, B-1050 Brussels, Belgium}

\author{Guido Caldarelli}
\affiliation{Istituto dei Sistemi Complessi CNR-ISC Rome, Italy}
\affiliation{DSMN and ECLT, Ca'Foscari University of Venice, Italy}
\affiliation{RARA Foundation - Sustainable Materials and Technologies, Venice, Italy}
\affiliation{London Institute for Mathematical Sciences, Royal Institution, London UK}

\author{Roberto Verdel}
\affiliation{The Abdus Salam International Centre for Theoretical Physics (ICTP), strada Costiera 11, 34151 Trieste, Italy}

\author{Marcello Dalmonte}
\affiliation{The Abdus Salam International Centre for Theoretical Physics (ICTP), strada Costiera 11, 34151 Trieste, Italy}

\begin{abstract}
Quantum computers and simulators offer unparalleled capabilities of probing quantum many-body states, by obtaining snapshots of the many-body wave function via collective projective measurements. The probability distribution obtained by such snapshots (which are fundamentally limited to a negligible fraction of the Hilbert space) is of fundamental importance to determine the power of quantum computations. However, its relation to many-body collective properties is poorly understood. 
Here, we develop a theoretical framework to link quantum phases of matter to their snapshots, based on a combination of data complexity and network theory analyses. 
The first step in our scheme consists of applying Occam's razor principle to quantum sampling: given snapshots of a wave function, we identify a minimal-complexity measurement basis by analyzing the information compressibility of snapshots over different measurement bases. 
The second step consists of analyzing arbitrary correlations using network theory, building a wave-function network from the minimal-complexity basis data.
This approach allows us to stochastically classify the output of quantum computers and simulations, with no assumptions on the underlying dynamics, and in a fully interpretable manner. 
We apply this method to quantum states of matter in one-dimensional translational invariant systems, where such classification is exhaustive, and where it reveals an interesting interplay between algorithmic and computational complexity for many-body states.
Our framework is of immediate experimental relevance, and can be further extended both in terms of more advanced network mathematics, including discrete homology, as well as in terms of applications to physical phenomena, such as time-dependent dynamics and gauge theories. 
\end{abstract}

\maketitle

\newpage

\section{Introduction}

Classifying phenomena is a long-standing goal of physics. Within the context of statistical mechanics, classifying phases has lead to profound insights on how collective properties of matter are determined. Seminal results include the classification of phases stemming from symmetry breaking~\cite{Sachdev_2023}, including roadblocks in lower dimensionalities~\cite{PhysRev.158.383, PhysRevLett.17.1133}, disordered~\cite{altland1997nonstandard} and critical systems~\cite{francesco2012conformal,caldarelli2007scale,barabasi2016network}, down to modern classification of quantum phases based on band structure properties and topological orders~\cite{ryu2010topological,10.1063/1.3149495, PhysRevB.83.075102,zeng2019quantum}. In the context of quantum matter, these schemes exploit wave-function knowledge, combined with specific inputs (e.g., dimensionality and symmetry), and connect those to observables, such as order parameters or transport properties, which are ideally suited for traditional solid state probes.

Over the last decade, synthetic quantum matter~\cite{gardiner2015quantum} - including trapped ions~\cite{blatt2012quantum,monroe2021programmable}, atoms in optical potentials~\cite{gross2017quantum,gross2021quantum,browaeys2020many}, and circuit QED~\cite{blais2021circuit,ripoll2022quantum} - has opened up a completely different scenario: thanks to many-body projective measurements, the whole wave function can actually be probed {\it directly}, albeit only {\it stochastically} and with a {\it limited} number of samples, due to the exponential growth of the Hilbert space preventing an exact tomography. The samples related to such global projective measurements are often referred to as {\it snapshots}.
Such capabilities have paved the way for research paths at the interface of quantum information and statistical physics: those include the capability of measuring nonlinear functionals of the density matrix via randomized measurements and shadow tomography~\cite{elben2023randomized,Chen2020Robust,vitalePRXQuantum2024,hu2025demonstration}, probing aspects of conformal field theories~\cite{sarma2025probing}, and the understanding of real-time dynamics from conditional probability distributions~\cite{ho2022exact,Claeys2022emergentquantum, Choi2023, PRXQuantum.4.010311, PRXQuantum.4.030322, PhysRevA.107.032215, PhysRevB.108.104317, mark2024maximum, Varikuti2024unravelingemergence, PRXQuantum.6.020343, PhysRevLett.134.180403, PhysRevB.111.144302}, just to name a few. However, in general, the connection between wave function snapshots and generic physical properties remains challenging, mostly due to the lack of a clear framework to understand such huge amount of information in a simple and interpretable manner, complementary to approaches based on neural network modelling~\cite{torlai2019integrating,torlai2018neural,Lange2023adaptivequantum,Miles2021, PhysRevLett.127.150504, Suresh_2025, moller2025learning}. 
This raises multiple questions: is it possible to connect the massive amount of information obtained in such devices to collective properties of quantum matter, in a manner that is at the same time generic and interpretable? Can we classify physical phenomena stochastically from limited available samples, or is prior knowledge of the setting fundamentally required to connect wave function samples to physical properties?

\begin{figure*}
    \centering
    \includegraphics[width=0.95\linewidth]{Figures/Algorithm.pdf}
    \caption{{\bf Schematic representation the wave function network analysis}: 
    \textbf{(a)} A given quantum state is produced either in experimental quantum simulations or in numerical experiments. \textbf{(b)} Snapshot datasets are built in different bases. This can be done experimentally via global projective measurements, or by following some sampling scheme in numerical simulations. 
    \textbf{(c)} Networks in different bases are built from datasets: each node corresponds to a snapshot and connections between nodes are established based on a cutoff of their affinity [see Sec.~\ref{sec:Construction_WFN}]. 
    Here we depict wave function networks (WFNs) obtained for the cluster Ising model [Eq.~\eqref{eq:ClusterIsing}], projected on the plane spanned by the two principal components of the data matrix. Color scale and nodes' sizes refer to their degrees.  Top panel is an Erdős–Rényi network (ERN) in the symmetry-protected topological phase, namely at $h=0.5$ in the $x$-basis.
    Bottom panel is a scale-free network (SFN) at the critical point $h=1.0$ in the minimal-complexity basis $y$. In the latter, we note the presence of highly connected nodes, called \emph{hubs}, characteristic of scale-free networks. Moreover, at this point two clusters start forming, due to the two-fold degeneracy of the ground state in the antiferromagnetic phase. Such clusters become more and more evident as $h$ increases. 
    \textbf{(d)} From the WFNs, the quantities of interest are computed, namely the intrinsic dimension $I_d$ and the degree distribution $P_k$. Top panel shows the $I_d$ of the cluster Ising model in the $x$-, $y$-, and $z$-basis as a function of $h$. Bottom panel shows the $P_k$ of the networks displayed in (c). Note that at the critical point, in the minimal-complexity basis $y$, the degree distribution is a power law, characterizing a scale-free network.}
    \label{fig:Algorithm}
\end{figure*}  

In this work, we propose a theoretical framework to characterize the stochastic sampling of wave functions, that leverages on and combines established tools in manifold learning and network theory. 
The rationale of our approach is that, in order to tame the large amount of correlations from many-body snapshots, it is of foremost importance to understand which basis direction can undergo a massive dimensional reduction - a version of Occam's razor principle -, and, once that `minimal-complexity' basis is identified, employ discrete, generic manifold characterization tools - in our case, network theory - to characterize correlations. 
We note that other network-theoretic approaches to quantum matter have been explored in the literature~\cite{llodrà2025unveilinghiddenfeatureskitaev, PhysRevLett.119.225301, PhysRevA.97.052320, 
Sokolov_Rossi_García-Pérez_Maniscalco_2022}, but they rely on different constructions.

Our pipeline is schematically illustrated in Fig.~\ref{fig:Algorithm}. Concretely, after collecting $N_s^{(\alpha)}$ samples in a suitable set of translational invariant measurement bases $\alpha$ (e.g., $\alpha = x,y,z$ for a spin model) - see Figs.~\ref{fig:Algorithm}(a) and~\ref{fig:Algorithm}(b) -, we embed them in distinct metric spaces, each one utilizing distances described by Parisi's two-replica overlap functions\footnote{We note that this corresponds to Hamming distance for spin-$1/2$ variables. We will comment on this specific choice and on its features (e.g., its relation to Euclidean distances) below. See Sec.~\ref{sec:Metric_Embedding} for more details. }~\cite{PhysRevLett.50.1946, mezard1987spin}. 

For each basis, we build a network representation of the samples: nodes represent snapshots, while links between nodes are determined in a deterministic manner based on node distance [see Fig.~\ref{fig:Algorithm}(c)]. The resulting networks are geometrical in nature, and we refer to those as \emph{wave function networks}~\cite{PhysRevX.14.021029}. Equipped with datasets and their network representations, we then proceed in a three-step analysis.

Firstly, we estimate the intrinsic dimension $I_d^{(\alpha)}$ of the corresponding data manifolds \cite{camastra2016intrinsic}, which, under rather generic assumptions, allows us to estimate the minimal number of degrees of freedom - a proxy for data compression and Kolmogorov complexity~\cite{PhysRevX.14.021029,che2025quantumcircuitcomplexityunsupervised} - required to capture the samples in each given basis. We identify the \emph{minimal-complexity basis} $\alpha^\text{min}$ as the one with the smallest intrinsic dimension $I_d^\text{min}$ among the ones we consider [see Fig.~\ref{fig:Algorithm}(d), upper panel].

Secondly, once the minimal-complexity basis is identified, we focus on the corresponding network. We characterize this via its degree distribution, which allows us to diagnose the presence or absence of correlations in data space in a practical manner [see Fig.~\ref{fig:Algorithm}(d), lower panel]. 

Finally, for networks showing random correlations (Erd\H{o}s-R\'enyi), we also investigate how $I_d^\text{min}$ varies as a function of real-space scale via a Kadanoff-type decimation scheme applied directly on the samples: this allows us to distinguish local from topological real-space correlations in a straightforward manner. 

The framework has three main advantages: it is highly interpretable (driven by the notion of classical state distance given by Parisi's overlap function), it is state agnostic (can be applied to any Hilbert space, dimensionality, etc.), and its applicability to analyze wave functions is scalable (typically a few thousand samples suffice for experimentally relevant system sizes). To demonstrate its use, we analyze one-dimensional (1D) translational invariant Hamiltonians, which are known to describe paramagnetic, critical (conformal), symmetry-broken, and symmetry-protected topological states of matter. We specifically focus on models which have been realized in Rydberg atom quantum simulators~\cite{de2019observation,ebadi2021quantum,guardado2021quench,schauss2018quantum, fang2024probing}, where many of the predictions we make are directly testable.

\begin{figure}
    \centering
    \includegraphics[width=\linewidth]{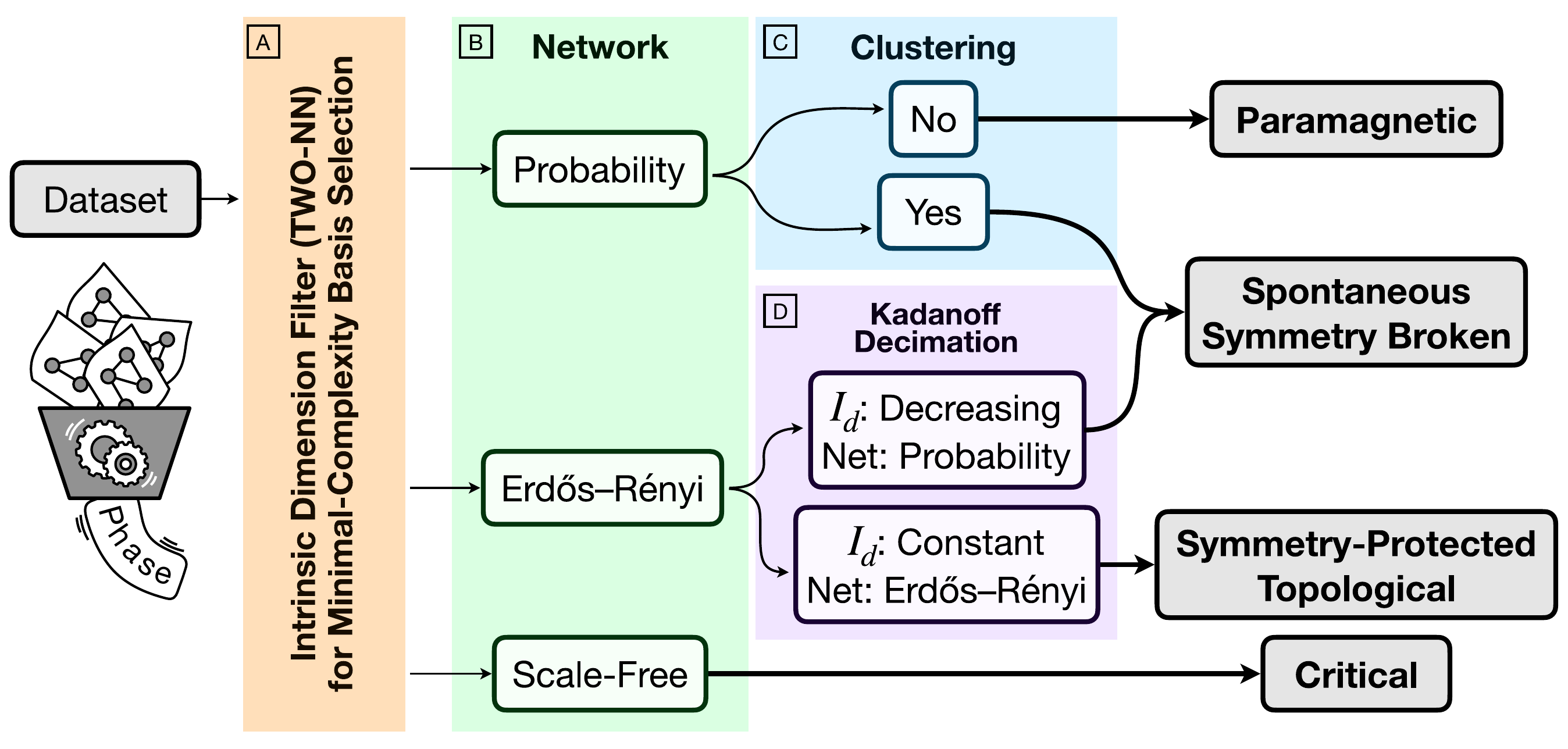}
    \caption{\textbf{Scheme of the stochastic classification.}
    After a filtering (a) through intrinsic dimension ($I_d$) to select the minimal-complexity basis, the network structure is studied (b). From top to bottom: Phases that show probability networks can be distinguished between paramagnetic and spontaneously symmetry-broken (SSB) ones by the presence of clusterization in data space (c), which reflects the ground state degeneracy typical of SSB phases. Erd\H{o}s-R\'enyi networks can signal either SSB or symmetry-protected topological phases. The two can be distinguished by computing $I_d$ after sampling on increasingly larger supports (d): In the former case, it decreases abruptly as soon as the support of the order parameter is reached, while in the latter it is left unchanged.
    Lastly, scale-free networks are associated to gapless phases and points}
    \label{fig:ClassScheme}
\end{figure}

Utilizing a combination of theoretical modeling and numerical simulations based on matrix product state (MPS) sampling, we show that the four states of matter available in one dimension - spontaneously symmetry-broken (SSB), paramagnetic, critical and symmetry-protected topological (SPT) phases - are all described by fundamentally distinct data structures. This classification is summarized schematically in Fig.~\ref{fig:ClassScheme}. Interestingly, our scheme treats on same footing gapped and gapless phases of matter, which is not usual in other contexts. 

Paramagnets feature small $I_d^\text{min}$ and a structured wave-function network, that recalls a probability distribution sampling - and thus, we refer to as probability network.
This is due to the product-state fixed-point description, which is reflected by presence of typical configurations (in the minimal complexity basis). The fluctuations around them are suppressed by the presence of a gap, causing accumulation on a small set of configurations. 
Critical matter display large $I_d^\text{min}$ and a scale-free network structure, where the degree distribution decays as a power of the degree. This is reflected in the presence of large fluctuations around typical configurations, for which the probability distribution does not concentrate on a small set of configurations.

Distinguishing symmetry-broken from SPT phases is instead more subtle. The key element there is to track how $I_d^\text{min}$ evolves under a Kadanoff-type decimation\footnote{Our scheme is directly inspired by traditional statistical mechanics~\cite{kadanoff1966}; it would be interesting to explore connections to recent developments in network theory~\cite{villegas2023laplacian}.}: for symmetry-broken phases, complexity shows a characteristic dependence on the decimation level, from large to small. This (abrupt) change corresponds to the decimation length at which symmetry-breaking takes place, that is, when it is as large as the support of the order parameter. The corresponding low-complexity network features a distinct structure, with prominent hubs (highly connected snapshots in our network construction), analogously to paramagnets.
For SPT phases, complexity is large and decimation-independent, and hence, the corresponding network structure is Erd\H{o}s-R\'enyi at all decimation scales, reflecting the inability of any local sampling to capture the order of the system.

The two most surprising observations of our framework are that:
\begin{enumerate}
    \item topological correlations are extremely simple to diagnose: they are epitomized by a minimal information compressibility at any scale, combined with an Erd\H{o}s-R\'enyi structure;
    \item critical points and phases are described by scale-free networks, signaling how universal behavior is not only carried out by low-order correlation functions, but is instead a characteristic of the entire sampled wave function. 
\end{enumerate}
These observations clarify that there is a deep, clearly interpretable link between the excessive amount of information contained in sampled wave function snapshots and collective phenomena. This connection might turn informative for a series of open challenges in quantum science, from the characterization of variational wave functions to quantum information tasks.

\section{Stochastic description of wave functions: from complexity to networks}\label{sec:Sec2}

In this section, we present the core theory at the basis of our stochastic classification of quantum matter. First, we describe the datasets of interest, then we introduce the main tools utilized for their analysis. 

\subsection{Datasets}\label{sec:datasets}

Our target datasets are collections of \emph{wave function snapshots}. These are the stochastic outcomes of projective measurements of local degrees of freedom in a many-body wave function, sampled according to Born's rule. For example, for a spin-1/2 system, a snapshot would be an array of $\pm1$ values, representing a spin configuration, whereas for a quantum gas microscope, it would be an array of local densities, representing a particle number configuration. 
This kind of data is routinely accessed in quantum simulation experiments, as well as sampled numerically, via, for example,  Monte Carlo or tensor network methods. Here we focus on spin-1/2 systems, with local Hilbert space dimension 2, but our reasoning may be applied to a vast plethora of experimental and numerical situations, not limited to the particular case study presented in this work. Extensions beyond what we discuss here include systems in the continuum and higher-dimensional local Hilbert spaces.

In contrast to classical systems, sampling the state of a quantum system involves a critical aspect, namely, the selection of the \emph{measurement basis}, i.e., the direction along which the local degrees of freedom are projected. To understand how crucial this aspect is, let us consider an elementary example. Measuring a spin-1/2 system in a ferromagnetic state along the $z$-direction, i.e., $\ket{\uparrow \dots \uparrow}$, one always gets the same spin configuration. However, along the $x$-direction, the snapshots would appear as random spin configurations, failing to provide information about the long-range order of the underlying quantum state. As this example illustrates, certain measurement bases - and consequently their associated datasets - provide a more relevant description of the state than others. 

It is thus clear that, in order to obtain a physically informative description of the state, while remaining agnostic about its properties, a set of appropriate orthogonal measurement directions must be considered, and a computable criterion to designate the most relevant one shall be defined. We thus examine datasets obtained by sampling spin configurations over a set of translational invariant measurement bases, defined by three orthogonal spin directions. In other words, we perform projective measurements where all spins are projected in the same direction $\alpha=x$, $y$, or $z$. We sample the snapshots using a perfect sampling algorithm for tensor networks simulations~\cite{Stoudenmire_2010, Ferris2012}, but, as noted above, these measurements are also easily accessible in modern experimental quantum platforms.

Let us consider a system with $L$ local degrees of freedom. We collect $N_s$ snapshots $\vec{x}^{(\alpha)}_s$, with $s=1,\dots,N_s$, in each of the considered measurement bases. In the case of a spin-1/2 system, a snapshot is denoted as $\vec{x}^{(\alpha)}_s = \{ \sigma^{(\alpha)}_{s,1}, \dots, \sigma^{(\alpha)}_{s,L}\}$, where $\sigma^{(\alpha)}_{s,i}$ is the value of the spin at site $i$ in the $s$th snapshot when measured along the $\alpha$ direction. The target datasets are then represented by matrices of dimension $N_s \times L$ [see Fig.~\ref{fig:Algorithm}(b)], that is,
\begin{equation}\label{eq:data sets}
    \mathbf{X}^{(\alpha)}=\left\{ \vec{x}^{(\alpha)}_1, \vec{x}^{(\alpha)}_2, \dots, \vec{x}^{(\alpha)}_{N_s} \right\}^\text{T}.
\end{equation}

Interpreting the measured values of the different local degrees of freedom as coordinates, the datasets above  can be geometrically regarded as  collections of $N_s$ points living on a $L$-dimensional data space.
At this stage, it is important to stress that, when the sampling probability is concentrated enough, which happens in some limiting cases, namely deep in ordered or paramagnetic phases, more samples of the same configuration could occur in the dataset. In our construction, those would appear as overlapping nodes on the same points of said $L$-dimensional data space, and we explain our way to treat them in the following sections. For future works, one could consider slightly different constructions, one example being weighted networks. A second example, that we will briefly treat later in App.~\ref{sec:Loop_nets}, is to consider sets of repeated configurations as single nodes with self-loops.

We note that typical values of $N_s$ for present-day quantum simulation experiments are such that $N_s \ll 2^L$ (hence, encompassing only a negligible fraction of the Hilbert space). Importantly, the methodology presented below has been devised to yield informative results when working in this regime.

\subsection{Wave function networks}\label{sec:WFN}

To characterize many-body quantum states, we apply the wave function network (WFN) description~\cite{PhysRevX.14.021029} to the fixed-basis datasets. A network~\cite{caldarelli2007scale, barabasi2016network}, as a concept borrowed from graph theory, consists of a set of nodes and links connecting pairs of nodes. The degree $k$ of a node is defined as the number of links connected to it, and the degrees in the network are distributed according to the degree distribution $P_k$. By analyzing the degree distribution, networks can be classified into different types. In the following, we briefly review the properties of three paradigmatic examples, which will play a key role below: Erdős–Rényi, scale-free, and structured networks.

Erdős–Rényi networks (ERNs) are characterized by a Poissonian degree distribution, which has a small variance around the average degree that sets the scale of the distribution. Such networks can be constructed by connecting pairs of nodes in a given set with a fixed probability, independent of the specific pair. For this reason, they are also referred to as random networks.

Scale-free networks (SFNs), first discovered in social structures ranging from authors collaborations networks to the world-wide-web, are a powerful structure describing a huge variety of phenomena  from diverse fields~\cite{barabasi2016network, caldarelli2007scale, newman2018networks}. They are characterized by a power-law degree distribution, $p(k)\propto k^{-\gamma}$, indicating large variability in node degrees. In particular, this reflects the presence of a great number of poorly connected nodes and of a few highly connected nodes called \textit{hubs}, which are instead absent in ERNs.

A third case we observe is constituted by what we refer to as \emph{probability networks}: those are characterized by a structured degree distribution showing a Poisson-like profile for low degrees, followed by a series of peaks at high connectivity. As it will be shown later, in these cases, the degree distribution, at large degree $k$, represents a histogram of the repetitions in the dataset, as same configuration can be accessed multiple times at finite lattice volume.

\begin{figure*}
    \centering
    \includegraphics[width=\linewidth]{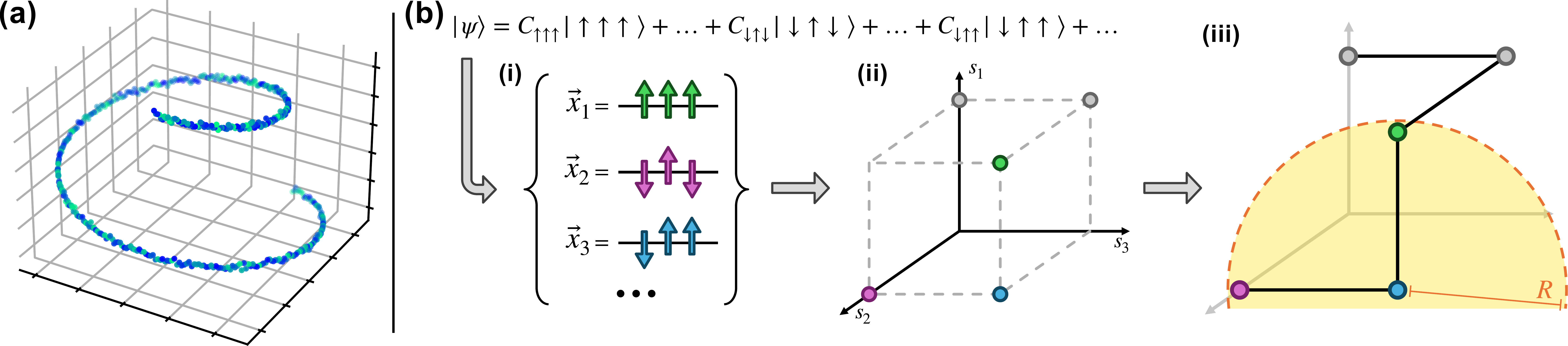}
    \caption{{\bf Schematic representation of intrinsic dimension and of the wave function networks.} \textbf{(a)} Data points in a high-dimensional data space are constrained by correlations to live on a lower-dimensional manifold, with minimal loss of information. In the toy example shown here, the dimension of the data space, i.e. the embedded dimension, is 3, while the intrinsic dimension is 1. \textbf{(b)} Construction of a wave function network: (i) Given a state $\ket{\Psi}$, snapshots are taken as explained in section \ref{sec:datasets}. (ii) Configurations in the dataset are points in the data space and represent the nodes of the networks. (iii) A metric is chosen on the data space, and a cutoff distance $R$ is fixed. Links are then drawn to connect pairs of nodes whose distance is below $R$. The basis label $\alpha$ has been dropped for the sake of clarity. 
    }
    \label{fig:DS_Tools}
\end{figure*}

Below we describe how a network description can be obtained from a collection of snapshots [see also Ref.~\cite{PhysRevX.14.021029}]. The procedure is presented schematically in Fig.~\ref{fig:DS_Tools}(b). We note that, in the classical limit, one recovers the construction of Ising nets reported in Ref.~\cite{sun2024network}.

\subsubsection{Nodes definition and embedding in metric spaces}\label{sec:Metric_Embedding}

WFNs are geometrical networks constructed from the fixed-basis datasets $\mathbf{X}^{(\alpha )}$ described in Eq.~\eqref{eq:data sets}. 
Their nodes are represented by the configurations $\vec{x}^{(\alpha)}_s$, which live on an $L$-dimensional data space. The latter can be equipped with an appropriate metric $d(\vec{x}^{(\alpha)}_r, \vec{x}^{(\alpha)}_s)$ to measure the affinity between configuration states in the $\alpha$-basis. 

For spin systems, a natural choice is to define this metric by means of the Parisi two-replica overlap between samples~\cite{PhysRevLett.50.1946}: \begin{equation}
    q\left(\vec{x}^{(\alpha)}_r, \vec{x}^{(\alpha)}_s\right) = \sum_i \sigma^{(\alpha)}_{r,i} \sigma^{(\alpha)}_{s,i},
\end{equation}
which defines the distance between samples $\vec{x}^{(\alpha)}_r$ and $\vec{x}^{(\alpha)}_s$ as~\cite{mezard1987spin}:
\begin{equation} 
    d(\vec{x}^{(\alpha)}_r, \vec{x}^{(\alpha)}_s) = \frac{ L - q\left( \vec{x}^{(\alpha)}_r, \vec{x}^{(\alpha)}_s \right)}{2}, 
\end{equation}
which, for spin-1/2 systems, corresponds to the Hamming distance:
\begin{equation}
    d(\vec{x}^{(\alpha)}_r, \vec{x}^{(\alpha)}_s) = \frac{1}{2} \sum_{i=1}^L \left| \sigma_{r,i}^{(\alpha)} - \sigma_{s,i}^{(\alpha)} \right|.
\end{equation}

We emphasize here that the specific reason why the two-replica overlap is chosen is that it is has tremendous success in linking collective state properties to physical phenomena in classical statistical physics. In the context of intrinsic dimension estimation, we note that such distances have been shown to display the same properties of Euclidean ones for different classes of spin models~\cite{TiagoPRX, TiagoPRXQuantum}.

\subsubsection{Intrinsic dimension}\label{sec:intrinsic_dimension} 

While in principle each wave function has multiple networks associated to it, each corresponding to different measurement bases, in Sec.~\ref{sec:datasets} we argued that some can reflect physical properties more directly than others. Identifying them is the first step of our analysis.

To identify the network that provides the most informative description of a state, we compare the measurement bases available and select the one whose associated dataset exhibits the lowest algorithmic complexity - a mathematical measure of the amount of information present in a sequence of symbols or data~\cite{li2008introduction}. We refer to this as the \emph{minimal-complexity basis}, denoted $\alpha^\text{min}$.
As a proxy for algorithmic complexity, we utilize the intrinsic dimension~\cite{camastra2016intrinsic, ghojogh2023elements}, that quantifies the minimal number of variables required to describe the dataset. We denote the intrinsic dimension of the dataset $\mathbf{X}^{(\alpha)}$ as $I_d^{(\alpha)}$. The minimal-complexity basis is therefore identified as the one whose dataset has the lowest intrinsic dimension $I_d^\text{min}$. For the case of spin $1/2$ systems, this reads:
\begin{equation}
    \alpha^\text{min}
    \,\,\Bigl|\,\,
    I_d^{(\alpha^\text{min})} = \underset{\alpha=x,y,z}{\mathrm{min}} I_d^{(\alpha)} \equiv I_d^\text{min}.
\end{equation}

The key point behind the concept of intrinsic dimension is that correlations between variables can sensibly constrain the structure of the data. From a geometrical viewpoint, this is reflected on the fact that data points typically lie on a lower-dimensional structure rather than on the full data space~\cite{wold1987principal, van2008visualizing}, in which such a structure is embedded (a sort of concentration of measure effect~\cite{bac2020local}). This is represented in Fig.~\ref{fig:DS_Tools}(a).
This choice is motivated by both physical and network reasoning. From the physics side, the intrinsic dimension very much relates to the common principle of effective field theory: bare equal observations, one prioritizes a theoretical description with as few variables as possible. 
From the network side, the intrinsic dimension is a good estimator for the Haussdorf dimension that on its own, for geometrical graphs, has a direct connection to the algorithmic complexity of the data description~\cite{staiger1993kolmogorov}.

While various methods exist to estimate the intrinsic dimension of data,  here we employ a local estimator known as the TWO-NN method~\cite{facco2017estimating}, which is based on statistics of nearest-neighbor distances between data points, under the  assumption of local data density homogeneity. 
We utilize this estimator for two reasons. First, it has already shown significant capability predictions in terms of physical phenomena~\cite{TiagoPRX, TiagoPRXQuantum, PhysRevB.108.134422, PhysRevE.109.034102,  PhysRevB.109.075152}. Second, although it requires key assumptions to be verified, its applicability regimes are well understood (in practice, data have to be described locally as convoluted Pareto distributions). It is worth noting that the algorithm strongly suffers from the appearance of repeated configurations in the dataset. We overcome this problem by appending a column of random numbers as an additional feature to the dataset, thus breaking the degeneracy between repeated configurations. The size of the support from which these numbers are uniformly sampled is directly proportional to the number of repetitions in the dataset, so that when none is present, the additional variable is uniformly set to zero. This procedure does not change the global picture nor quantitatively spoils the physical content of the data. We note that it is also possible to extend our analysis utilizing estimators for discrete metric spaces~\cite{Mannila, PhysRevLett.130.067401, Acevedo2025}, as well as other proxies of data compressibility~\cite{chen2025zippingmanybodyquantumstates, PhysRevX.9.011031}. We leave that exploration to future works. 

Because the intrinsic dimension is computed from the dataset alone, it can be regarded as a property of the network depending only on its node set. Additional information, however, can be extracted by looking at the full network structure. The next paragraph is devoted to this goal.

\subsubsection{Network construction and structural analysis}
\label{sec:Construction_WFN}

The last step towards the network construction is the construction of links. As anticipated, we follow a geometrical procedure, i.e., we activate links between pairs of data points whose distance is below a chosen cutoff radius $R$. In this work, our standard choice for this cutoff is the average $n^\text{th}$-nearest neighbor distance computed on the dataset:
\begin{equation}\label{eq:cutoff_def}
    R = \overline{d_n} =
    \frac{1}{N_s} \sum_{s=1}^{N_s} d(\vec{x}^{(\alpha)}_s, \vec{x}^{(\alpha)}_r) \,\Big|_{r=n\text{-th neighbor of }s},     
\end{equation} 
where the ordering of neighbors is induced by the metric embedding. We note that the qualitative features of the network are robust to slight modifications of the cutoff, as we illustrate in some specific cases below.

When repeated configurations are present in the dataset, the sum in Eq.~\eqref{eq:cutoff_def} is dominated by zeros. Because of this the cutoff distance may become smaller than 1. In this case, the network connects only overlapping nodes, representing equal configurations. Thus, the degree of each node is the number of its copies, and the degree distribution represents an histogram of the repetitions of each node. It is for this reason that we call this kind of networks probability networks.

Given the datasets $\mathbf{X}^{(\alpha)}$ associated to the wave-function of a quantum state, we thus construct different networks. As mentioned above, the intrinsic dimension allows us to select a measurement basis where information is more structured. Consequently, we can analyze properties of the WFN in the minimal-complexity basis that depend on the information carried on the edges, thereby revealing the imprint of correlations in a way specific of our network construction.  

To this goal, several properties of the network can be studied. In this work, we focus on the degree distribution $P_k$. We note here that this kind of analysis has been already employed in Ref.~\cite{PhysRevX.14.021029} to characterize the output of a Rydberg quantum simulator. However, it is important to notice that the analysis done in the aforementioned work, was limited to snapshots taken in a single measurement basis (similar to classical statistical mechanics~\cite{sun2024network}). While such approach is indeed useful to characterize the system dynamics, it is clearly not sufficient for classifying states of matter, where information about multiple measurement bases is fundamental.

It is important to anticipate that deeper levels of classification are also accessible within our formulation. Indeed, the \emph{dimension} and \emph{connectivity} of the network are just some of the most direct properties that can be probed: deeper knowledge is accessible via more complicated geometrical~\cite{bronstein2021geometricdeeplearninggrids} and topological tools~\cite{sun2024network}, such as persistent homology~\cite{wasserman}, just to name a widespread approach.

\section{Stochastic classification of phases of matter in one dimension}\label{sec:Classification}

In one-dimensional translationally invariant spin models, the nature of correlations allow to distinguish between different phases of matter. These fall into four categories: ordered (e.g., ferromagnets or antiferromagnets), disordered (paramagnets), critical (or conformal), and symmetry-protected topological phases. 
Importantly, for a proper phase identification, one needs to know {\it a priori} what physical observables are relevant to the description of the phase itself, or at least, to input the symmetry content of the Hamiltonian dynamics that realizes such phases as ground states.

Here we illustrate our network theory-based classification of phases of matter, which does not rely on any prior knowledge. The process involves a first step, of identifying the minimal-complexity basis by computing $I_d$, followed by the analysis of the corresponding WFNs. To each possible phase of matter observable in one-dimensional spin-1/2 systems, we associate a network structure that characterizes the corresponding wave function, and the results are illustrated in the scheme in Fig.~\ref{fig:ClassScheme}. All the predictions presented in this section will be supported by observations in Sec.~\ref{sec:Results}.
We emphasize that this classification is complementary (in a manner that we specify below), both at the mathematical and at the physical level, to previous classifications of quantum matter. At the mathematical level, we leverage on a stochastic description, which naturally calls for very different tools with respect to algebraic and group-theoretic ones. At the physical level, our ultimate goal is to classify phases realized in quantum processors and simulators, where exact global symmetries might be spoiled by dissipation, but full wave function properties are available. As noted above, further layers of classification might be possible, based on, e.g., topological data analysis. We leave the exploration of those tools to future work.

\subsection{Paramagnetic phases: low complexity, structured networks}

In paramagnetic phases, all correlation functions are expected to decay exponentially as a function of distance. The fixed point description is that of a product state, and no symmetry is broken. Information can be compressed efficiently, and as such, the minimal complexity basis will feature small intrinsic dimension
and probability networks, as defined in Sec.~\ref{sec:Construction_WFN}.

For the sake of clarity, we illustrate the corresponding generic network properties via an example (partly recalled already in Sec.~\ref{sec:datasets}): a trivial paramagnet. In that case, spins align along the direction of the external magnetic field, say $x$. 
Sampling along $y,z$ will produce random samples, evenly filling the configuration space, resulting in a random network with very large intrinsic dimension; sampling along $x$ will instead produce a dataset dominated by repeated samples of typical configurations, and of its neighbors (in terms of their Hamming distance). As described in the previous sections, this leads to the emergence of probability networks. The nodes' degrees introduce a hierarchical structure led by the typical configuration, reminiscent of the fixed point (classical) description, followed first by neighboring configurations, and then by farther ones, which are sampled with smaller and smaller probabilities.

To understand what we mean by ``small intrinsic dimension'', take the extreme case of an infinite magnetic field. Samples along the $x$-direction will all look the same, hence, knowing the value of the first spin is enough to determine the full configuration. Considering a system with a bigger size does not change the complexity of the string, thus $I_d\sim O(1)$ with respect to the size $L$. A finite but big magnetic fields introduces perturbations that modify weakly that behavior. This qualitative picture elucidates how $I_d$ captures the simplicity of a paramagnetic state. Crucially, $I_d$ enters our analysis only as a filtering criterion for selecting the minimal-complexity basis, ensuring that its qualitative character does not influence the final results.

\subsection{Spontaneous symmetry-broken phases: low complexity, structured networks, and real-space Kadanoff decimation}

Ordered phases of matter are characterized by an order parameter describing the corresponding spontaneous symmetry breaking. Their sampling structure will depend drastically on the relation between the support of the order parameter, that we denote as $K$, and the sampling strategy.  

\paragraph*{$K=1$ case. -}We start from the case where the order parameter has support on one site, and the sampling is single-site resolved. In this case, the minimal-complexity basis will correspond to that of the order parameter (i.e., the eigenbasis of the operator associated to the order parameter), or, in case the latter is not parallel to any of the considered directions, to the basis with which has the largest overlap. 
Consequently, following a similar argument to that of disordered phases, the complexity of datasets resulting from sampling the ground states of ordered phases must be very low, e.g., $I_d=O(1)$ in the fixed point limit. We argue that this behavior is connected to the presence of a local order parameter, namely, if the phase can be described by a local order parameter, then there exists a local basis in which $I_d^{(\alpha)}$ is naturally minimal.

Differently from disordered phases, though, this manifold will also exhibit clustering, since the spontaneous symmetry breaking of discrete symmetries (the only one possible in 1D quantum models) will lead to degenerate ground states that are - in terms of the Hamming distance - very far apart. 
We notice that this conclusion is a direct consequence of Mermin-Wagner-Hohenberg theorem~\cite{PhysRev.158.383, PhysRevLett.17.1133}, since, for continuous symmetries, it is not clear if such clustering would necessarily occur.

\paragraph*{$K>1$ case. -} We now turn to the case of order parameters with support beyond one lattice site (dimer, and in general nematic phases). In this case, sampling over a single site is uninformative about the state. Utilizing a MPS representation, this can be noted by the fact that an ordered phase imposes a symmetry $\mathcal{G}$ over multiple sites, and local sampling is necessarily affected by the residual entropy obtained while tracing a single site. Such an entropy generation mechanism is akin to that found in computing minimal entropies in symmetric MPS~\cite{li2024symmetryenforcedminimalentanglementcorrelation}.

What genuinely characterize ordered phases is the scaling of complexity as a function of $\ell$, that we denote as $I_d(\ell)$, where $\ell$ is the scale at which the system is probed. This scaling can be obtained from the original data by performing a coarse graining procedure akin to Kadanoff decimation [see Fig.~\ref{fig:MPS-scheme}(a)]: given a system of length $L$, at each step, we combine sites into pairs, and define new variables spanning the same configuration space as the original ones. For instance, in the case of spin-1/2, the first step of coarse graining takes
\begin{equation}
    X^{(1,\alpha)} = \{\sigma_1^\alpha, \dots,\sigma_L^\alpha\},
\end{equation}
onto
\begin{equation}
    X^{(1,\alpha)}_2 = \{(\sigma_1^\alpha \cdot\sigma^\alpha_2), (\sigma_3^\alpha \cdot \sigma^\alpha_4),\dots,(\sigma_{L-1}^\alpha \cdot\sigma^\alpha_L)\}.
\end{equation}

From the point of view of complexity, one can then clearly separate two regimes: 
\begin{enumerate}
    \item $2^{\ell} < K$: in this regime, complexity is large, as it is impossible to faithfully represent dimensional reduction due to residual entropy;
    \item $2^{\ell} \geq K$: after the sampling scale exceeds $K$, the situation is akin to that of a $K=1$ phase, and thus the data display minimal $I_d$.
\end{enumerate}
As we will discuss below in Sec.~\ref{sec:kadanoff}, this complexity transition as a function of $\ell$ 
will be key in discerning the behavior of ordered versus topologically ordered phases. 

Summarizing, we have that ordered phases can be diagnosed by either: 
\begin{itemize}
    \item A structured network with low intrinsic dimension and clustering ($K=1$),
    \item A series of coarse-grained ERNs at every decimation scale, whose intrinsic dimension abruptly diminishes when the coarse-grain scale is as large as the support of the order parameter, and that ultimately leads to a structured network with low intrinsic dimension ($K>1$).
\end{itemize}

\subsection{Symmetry-protected topological phases: high complexity, Erd\H{o}s-R\'enyi networks}

In SPT phases, all correlations decay exponentially as a function of distance~\cite{zeng2019quantum}, and there is no local order parameter. Sampling the wave-function in any direction will thus return a random Erdős–Rényi network: distinct regions of space are sampled independently and, at the level of single spins, are uncorrelated. 

However, the existence of a non-local order parameter fundamentally affects the network complexity. At odds with paramagnetic and ordered phases, here, it is not possible to find a direction where complexity is of order $O(1)$. This implies that the intrinsic dimension is expected to increase with system size: in fact, one could see the sampling of a SPT state as a collection of independently sampled regions where, however, no single direction is preferred. This can be connected to the fact that, under local sampling, it is not possible to deduce the order parameter of a SPT phase (which is non-local), as its fixed description is always a MPS with finite bond dimension across the lattice~\cite{zeng2019quantum}. 

\subsection{Conformal phases and critical points: high complexity, scale-free networks}

At conformal critical points, there exists a set of correlation functions that decays algebraically with distance. One basic aspect of such critical points is that their wave function snapshots can be interpreted as sampling single slices of a isotropic path integral, i.e., where space and imaginary time direction are both governed by the same dominant correlations. The corresponding network structure is expected to show no scale, and, in particular, no scale at the level of degree distributions. We thus expect a description in terms of scale-free networks, at least along one measurement axis. The scaling of the intrinsic dimension can then be inferred from the known properties of classical partition functions at conformal criticality~\cite{TiagoPRX}: the intrinsic dimension will be large and increase with system size according to scaling laws, at least, for sufficiently large sampling. In summary, sampling conformal field theories will return scale-free networks with large intrinsic dimension.

\subsection{Sorting out symmetry breaking from symmetry-protected topological phases: more on Kadanoff decimation and complexity}
\label{sec:kadanoff}

The formulation above provides a view on the network structure of gapped and gapless phases. However, in order to formulate a clear criterion to distinguish those, it is important to better describe the renormalization group decimation scheme that is formulated at fixed embedding dimension, which was outlined when describing SSB phases. This is necessary not only to limit unavoidable finite size effects, but also to clarify what a ``topological'' order parameter can be in the data structure of SPT phases, which can clarify how those can be interpreted as phases where a hidden, non-local symmetry has been spontaneously broken.

\begin{figure*}
    \centering
    \includegraphics[width=\linewidth]{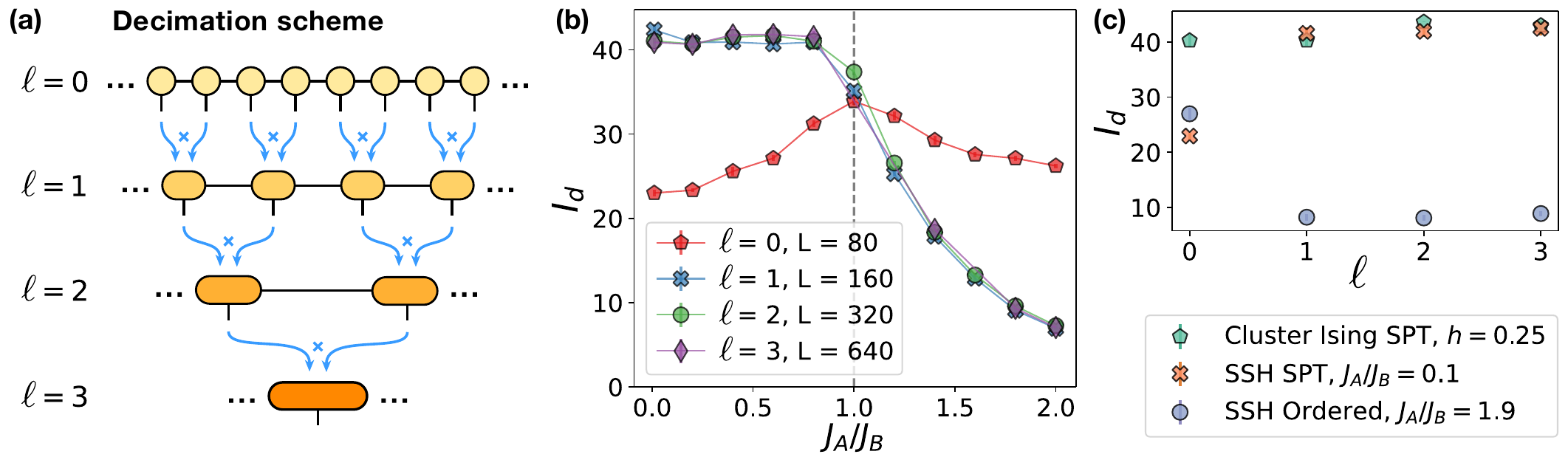}
    \caption{{\bf Connection between sampling support and complexity}.
    \textbf{(a)}  Decimated sampling procedure: pairs of neighboring sites in the MPS are merged and sampled with respect to their product as spin-1/2 variables. Performing this procedure $\ell$ times is equivalent to sampling observables with support on $2^\ell$ sites. To compare results for different values of $\ell$, we consider systems of different sizes $L$ so that $L/\ell$ is constant. \textbf{(b)} $I_d$ of SSH model [Eq.~\eqref{eq:SSHmodel}] for samplings with different supports. In the dimerized ordered phase, $J_A/J_B>1$, the complexity drops significantly once the support of the sampling exceeds 2. On the other hand, it doesn't decrease in the SPT phase, $J_A/J_B<1$. Note that the sampling with support bigger than one site breaks the $U(1)$ invariance of the model, while preserving the $\mathbb{Z}_2$ symmetry. For this reason, as some constraints are dropped, the complexity of the state increases from $\ell=0$ to $\ell=1$. \textbf{(c)} $I_d$ as a function of $\ell$. In the dimerized ordered phase of SSH model, as soon as the support of the sampling exceeds 2, the $I_d$ drops significantly. On the other hand, in the SPT phases of both Cluster Ising and SSH models, shown here respectively at $h=0.25$ and $J_A/J_B=0.1$, the $I_d$ never decreases, as there's no way to reduce the complexity of the state by sampling with respect to non-global operators.}
    \label{fig:MPS-scheme}
\end{figure*}

We propose thus to follow a decimation procedure that, starting from level $\ell=0$ (no decimation) and lattice volume $L$, analyses successive steps applying iteratively a parity blocking of size 2. At each step, the analysis is performed with the same number of samples, but at system size $L'=L\times2^\ell$. This allows to compare samplings, and thus $I_d$, at fixed embedding dimension $L'/2^\ell= L$. The procedure is schematically illustrated in Fig.~\ref{fig:MPS-scheme}(a).

Under this decimation procedure, we expect {\it both} SPT and ordered phases to display a characteristic dependence on $\ell$. The main difference will be on magnitude and scaling with $\ell$: For a SPT phase, $I_d$ as a function of $\ell$ shall be large, and  roughly independent of $\ell$ itself; oppositely, complexity for a SSB phase shall decrease once we hit the decimation scheme $\ell_*$ for which the total size of the blocking comprises the order parameter support, i.e., $2^{\ell_*}\geq K$.

The discussion above emphasizes how, at the level of sampling, there are fundamental differences between ordered and SPT phases, that, however, require considering properly spanned basis to be identified and highlighted. 
These basic differences can be traced to the capability of the network to be sensitive to local and non-local order parameters in very different manners. We note that such difference has also be connected to other information-based properties, like entanglement length~\cite{PhysRevA.71.042306}, which is connected to our circuit reasoning via localizable entanglement. 

Most importantly, the discussion above proposes a basic connection between the computational (circuit) complexity of a given MPS, and the minimal algorithmic complexity of its sampling: at fixed entanglement, states that are harder to prepare (SPT) necessarily feature much larger algorithmic complexity when compared to states that are easier to prepare (ordered phases and paramagnets). Such a fundamental difference might however require sampling of larger unit cells, depending on the type of ordered phase.

\paragraph*{Illustration: the Su-Schrieffer-Heeger model. - }For the sake of clarity, we now illustrate the predictions of our theory above in the context of the Su-Schrieffer-Heeger (SSH) model~\cite{PhysRevLett.42.1698}. The latter is a model describing fermions on a lattice, whose Hamiltonian reads:
\begin{equation}\label{eq:SSHmodel}
    H = J_A \sum_{j=1}^{L/2} a^\dagger_{2j-1} a_{2j} +  J_B \sum_{j=1}^{L/2-1} a^\dagger_{2j} a_{2j+1} +\text{h.c.}
\end{equation}
where $a^\dagger_j (a_j)$ are creation (annihilation) operators on a chain comprising $L$ sites. The model hosts a phase transition at $J_A=J_B$, separating a SPT phase ($J_A/J_B<1$) from an ordered phase ($J_A/J_B>1$). The latter is a valence bond solid (VBS) with unit cell of 2 sites, so in this case, $\ell_*=1$.
The states are sampled along a single measurement basis, namely the occupation basis. For decimation schemes $\ell\geq1$, the sampled value corresponds to the parity of the total occupation number on the sampling support $2^\ell$, i.e., the sum of the occupation numbers modulo 2.

Our theory predicts that, for single site sampling, both phases will feature $I_d\gg 1$. Instead, for $\ell\geq1$ sampling, while the SPT phase will still feature large complexity, the VBS solid phase will have a finite $I_d$, approaching 1 in the limit of $J_B/J_A\ll 1$. 
The network structure will be ER for all $\ell$ in the SPT phase, and for $\ell<1$ in the VBS phase, where it will become a probability network for $\ell\geq1$. At the critical point, as usual, we will have a SF network.

We test these predictions numerically, and the results shown in Figs.~\ref{fig:MPS-scheme}(b) and ~\ref{fig:MPS-scheme}(c) confirm them.  Additionally, we also notice that $I_d$ grows in the SPT phase after the first decimation step. We argue this is due to the decimation scheme breaking the $U(1)$ invariance of the model: Indeed, in general, symmetries introduce constraints in data space that, in this case, lower $I_d$ at $\ell=0$ with respect to $\ell\geq1$, where the constraint is removed. To support this argument, in Fig.~\ref{fig:MPS-scheme}(c), we present also the data from the SPT phase of Cluster Ising model [see Sec.~\ref{sec:Models}], where this doesn't happen, showing that $I_d$ remains constant under decimation in the SPT phase.

In Fig.~\ref{fig:MPS-scheme}(b) the behavior of $I_d$ at decimation scale $\ell$ is reported. Consistently with the expectations from Sec.\ref{sec:Classification} a decimation scheme with $\ell>\ell_*$ yields a larger $I_d$ for the SPT phase  while it yields a lower $I_d$ in the VBS phase. In Fig.~\ref{fig:MPS-scheme}(c), the scaling behavior of $I_d$ for ordered SPT and ordered phases is presented. In such figure we report also data for the SPT phase of the cluster Ising model, whose magnitude remains unaffected due to the large support of the order parameter, which is extended over the whole system.

\subsection{Relation to other classification schemes}

Here, we briefly discuss how the classification presented in this work compares with others, both in terms of approach and applicability regimes. We refer to already used classification schemes based on symmetry-breaking ({\it \`{a} la} Ginzburg-Landau), topological insulators and superconductors and SPT phases~\cite{altland1997nonstandard,10.1063/1.3149495,zeng2019quantum}, and unitary conformal field theories~\cite{francesco2012conformal}. 
\begin{itemize}
    \item Possibly the main difference between our scheme and those above is that our classification does not posit (neither is informative about) spectral properties. In particular, it treats on same footing gapless and gapped matter, and, within the latter, topological and SSB phases. The capability of distinguishing between those solely based on wave function properties is in fact a key outcome of our approach;
    \item Another key difference is the relation between phases and complexity. For instance, the classification of SPT phases, as well as that of conformal field theories (CFTs), make key connections between computational complexities in the context of tensor network, and phase properties: notable examples are the relation between the central charge and entanglement, and the degeneracy of the entanglement spectrum and edge modes~\cite{amico2008entanglement, Calabrese_2009}. Our approach is only indirectly sensitive to computational complexity; instead, it relies heavily on algorithmic complexity as a guiding principle. The difference between the present and other approaches in relating complexity to physical properties could provide an whole new angle on understanding the computational versus algorithm complexity dichotomy in the context of many-body states (which are {\it per se} highly constrained classes of probability distributions);
    \item It is not immediately clear how to compare classification schemes, due to the fact that typically priors are needed (e.g., the presence of a gap). Modulo this important caveat, the present depth of the scheme we propose is surely more limited with respect to the others, in particular with respect to CFTs and SPT phases. However, it is important to note that, even in those fields, full classifications have required subsequently increasing level of sophisticated mathematics to be uncovered. We expect the same to be true in the context of stochastic classification: so far, we have utilized only basic manifold properties, such as dimensionality and connectivity. More advanced tools, such as persistent homology or the study of higher order simplices, would likely be fundamental to further refine our approach;
    \item Another important difference is in how
    different classifications utilize renormalization group (RG) concepts. In particular, and somewhat surprisingly, RG decimation is only required here to distinguish massive phases, while in traditional classification schemes, RG is often associated to flow to critical points. However, a recent work has related some network properties to critical exponents in classical statistical mechanics~\cite{brunner2025snapshotrenormalizationgroupquantum}, which could be another extremely promising direction both in terms of refining classification, and in terms of connecting it to RG flows also for critical matter; 
    \item Finally, the main difference - and in fact, our main motivation - is that our scheme builds on the absence of any prior, including any underlying microscopic symmetry. This is particularly important for synthetic quantum matter, where microscopic symmetries are essentially absent on experimentally relevant timescales due to (possibly small, but always finite) dissipation rates. 
\end{itemize}

\section{Numerical Results}\label{sec:Results}

\subsection{Lattice Models}\label{sec:Models}
In order to illustrate and scrutinize the predictive power of our theory, in addition to the SSH model introduced in the previous section, 
we discuss three prototypical examples: the Ising model, the Cluster-Ising model and the XXZ model. The rationale behind comparing these spin chains is that they display different phases where the characterization power of our method can be best showcased, according to what was anticipated in the previous section.\\

\subsubsection{Ising Model}
The transverse field Ising model reads:
\begin{equation}\label{eq:Ising}
    H=-\sum_i \sigma^z_i \sigma^z_{i+1} -h \sum_i \sigma^x_i,
\end{equation}
where $\sigma^{\alpha}_i, \alpha=x,y,z$ are Pauli matrices. At the point $h=1$ the system undergoes a second-order quantum phase transition from a quantum paramagnetic (PM) phase for $h>1$ to a ferromagnetic (FM) phase for $h<1$. The phase diagram is drawn in Fig.~\ref{fig:Results_ID}(d) as a function of the transverse field $h$.

\subsubsection{Cluster-Ising Model}
The Cluster-Ising model, as described in \cite{SmacchiaClusterIsing2011, Son_2011}, exhibits a cluster-like interaction competing with an Ising-like antiferromagnetic interaction. Remarkably, the model undergoes a quantum phase transition between an Ising phase with a non-zero magnetization and a cluster phase that can be characterized by a string order parameter.
The Hamiltonian of the model reads:
\begin{equation}\label{eq:ClusterIsing}
    H=-\sum_i \sigma^x_i \sigma^z_{i+1}\sigma^x_{i+2} +h \sum_i \sigma^y_i\sigma^y_{i+1},
\end{equation}
where $\sigma^{\alpha}_i, \alpha=x,y,z$ are Pauli matrices.
The quantum critical point appears at $h=1$. 
For $h<1$, there is a symmetry-protected topologically (SPT) ordered phase, protected by the symmetry $\mathds{Z}_2 \times \mathds{Z}_2$; for $h>1$, the system ground state is reminiscent of an Ising antiferromagnet (AMF). The phase diagram of the Cluster-Ising model is represented in the red region of Fig.~\ref{fig:Results_ID}(d) as a function of the parameter $h$ that drives the transition.

\subsubsection{XXZ Model}
The spin $S=1/2$ XXZ model~\cite{Fradkin_2013, Franchini_2017} is a spin chain with uniaxial anisotropy, whose Hamiltonian reads:
\begin{equation}\label{eq:XXZ}
    H=J\sum_i \left( \sigma^x_i \sigma^x_{i+1} +\sigma^y_i \sigma^y_{i+1} \right)+J_z\sum_i \sigma^z_i \sigma^z_{i+1}.
\end{equation}
For $J<0$, ferromagnetic order is preferred along the x-y plane, while when $J>0$ we have an antiferromagnet in the plane. 
The strength of the uniaxial anisotropy along the $z$ direction is set by the ratio $J_z/J$. For $|J_z/J|<1$ the chain remains critical. The ground state has zero magnetization and it is a paramagnet characterized as a spinon vacuum. 
At $J_z/J=-1$ and $J_z/J=1$ there are two quantum critical points, the second one is particularly interesting as it features a Berezinsky–Kosterlitz–Thouless (BKT) type of transition. In the region $J_z/J<-1$ the ground state is ferromagnetic along the $z$ direction, while for $J_z/J>1$ the preferred order is antiferromagnetic. For the rest of this work, we will set $J=1$. This phase diagram is represented in Fig.~\ref{fig:Results_ID}(d) in green as a function of the parameter $J_z$.

\subsection{Intrinsic dimension}

\begin{figure*}
    \centering
    \includegraphics[width=\linewidth]{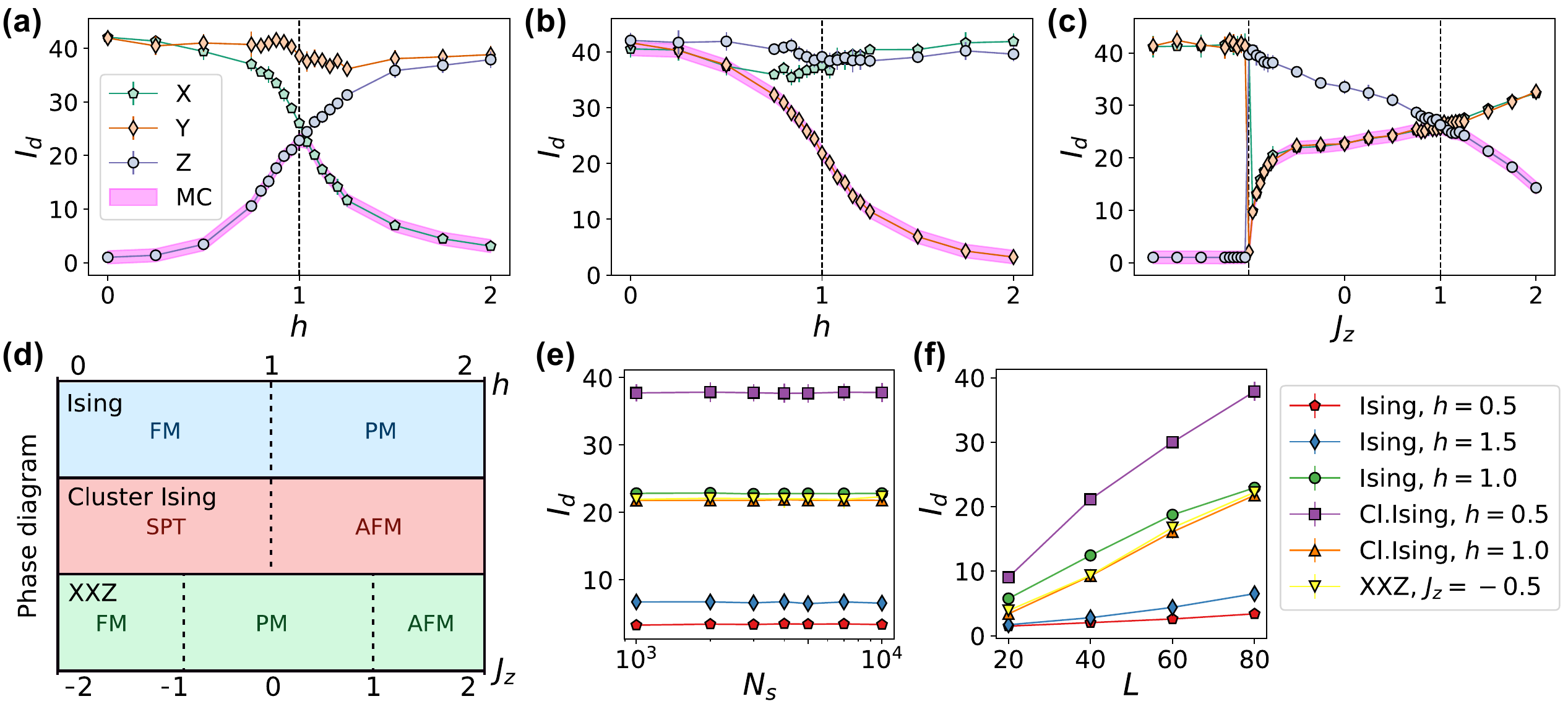}
    \caption{{\bf Intrinsic dimension of the ground states of 1D quantum spin model.} \textbf{(a-c)} Intrinsic dimension of the ground states of the 1D quantum spin chains under examination, with $L=80$, sampled in $x$, $y$ and $z$ bases. We observe low value of the $I_d$ in the minimal-complexity basis (highlighted in pink) deep in the ferromagnetic and paramagnetic phases, and high values in SPT phases and inside critical phases and points. All values are computed for a number of samples $N_s=10^3$. \textbf{(d)} Phase diagrams of the three considered models. \textbf{(e)} Scaling of $I_d$ with number of samples $N_s$. The computed values are stable if $N_s$ is increased by one order of magnitude. \textbf{(f)} Scaling of $I_d$ with system size $L$. In ordered and disordered phases, $I_d$ has a weak dependence on $L$. In critical and SPT phases, the $I_d$ grows faster with $L$. }
\label{fig:Results_ID}
\end{figure*}

In this section we show the results obtained for the intrinsic dimension. All configurations that make up the datasets are sampled via perfect sampling~\cite{Stoudenmire_2010, Ferris2012} from ground states obtained from tensor network density matrix renormalization group (DMRG) simulations~\cite{White_1993, Schollwock_2005, schollwock2011density}. As already stated, the estimator chosen for $I_d$ is the TWO-NN algorithm, which has been applied on datasets of $N_s=10^3$ samples for each value of the control parameter and for every choice of the basis. Repetitions occurring in SSB and disordered phases are treated as explained in Sec.~\ref{sec:intrinsic_dimension}.

The intrinsic dimension of the quantum Ising model ground state is shown in Fig.~\ref{fig:Results_ID}(a). In the $z$-basis, $I_d$ is low in the ferromagnetic phase and grows while crossing towards the paramagnetic phase. Indeed, the extreme cases are $h=0$ on one side, where the only sampled configuration is the fully polarized state, and $h\to\infty$ on the other side, when the dataset looks random. Analogously, $I_d$ behaves in the opposite way when the state is sampled in the $x$-basis. Finally, the algorithmic complexity saturates everywhere in the $y$-basis, in accordance with the fact that such a dataset carries no information about the physics of the system. Thus we identify as minimal complexity basis the $z$-basis in the ferromagnetic phase and the $x$-basis in the paramagnetic phase. 
Note that in the critical region, the data of both $x$ and $z$ bases feature a similar algorithmic complexity, as a consequence of the model's self duality. Hence, at the critical point, any of these bases can be chosen as the minimal-complexity basis.

The results for the cluster Ising model are portrayed in Fig.~\ref{fig:Results_ID}(b). According to the classification presented in Sec.~\ref{sec:Classification}, $I_d$ is high in all bases in the SPT phase, as no local sampling would be able to significantly lower the complexity. Nevertheless, it decreases when transitioning into the antiferromagnetic phase when observed in the $y$-basis. The latter is identified as minimal-complexity basis in the whole phase diagram.

Finally, the XXZ model is shown in Fig.~\ref{fig:Results_ID}(c). Due to the symmetry on the $xy$-plane, the behavior of $I_d$ in the $y$ basis mimics that of the $x$-basis throughout the phase diagram. Hence, in the following, we only compare the $x$ and $z$ bases.
We observe a low intrinsic dimension along $z$ and a high one along $x$ for $J_z<-1$ and $J_z>1$, compatible with the presence of, respectively, a ferromagnetic and an antiferromagnetic state along the $z$ direction. The $z$-basis is then the minimal-complexity basis in these two phases. Conversely, for $|J_z|<1$, where the ground state features long range correlations, $I_d$ is high both in the $z$-basis and in the minimal-complexity basis $x$. 
Notably, the behaviors at the two transition points are different. For $J_z=-1$, $I_d$ features a jump, compatible with a first order transition. On the other hand, at $J_z=1$ there is a smooth crossing between the increasing $I_d^{(x)}$ and $I_d^{(z)}$, which decreases slowly. This is reminiscent of the nature of the transition. Indeed, in a BKT transition such as the one at $J_z=1$ we expect the correlation to fade slowly compared to first- and second-order phase transitions.

Fig.~\ref{fig:Results_ID}(e) shows the scaling of the scaling of $I_d$ in the minimal-complexity basis with the number of samples $N_s$. We can observe how it remains extremely stable as the number of samples are increased up to one order of magnitude more. Lastly, in Fig.~\ref{fig:Results_ID}(f) we present the scaling of the minimal-complexity with system size. As expected, in ferromagnetic and paramagnetic phases it grows extremely slow with system size in contrast to critical points and phases and SPT phases. Thus, a rough classification can already be performed solely on the basis of the algorithmic complexity in the minimal-complexity basis, which is low in ordered and paramagnetic phases, while it is high in SPT and critical phases. We expect the first two to be distinguishable via a cluster analysis [see Fig.~\ref{fig:Algorithm}(c) for an intuitive visualization], while the network analysis will allow to distinguish the last two, as explained in the next section.

\subsection{Network Analysis}
\label{sec:Net_analysis}

\begin{figure*}
    \centering
    \includegraphics[width=\linewidth]{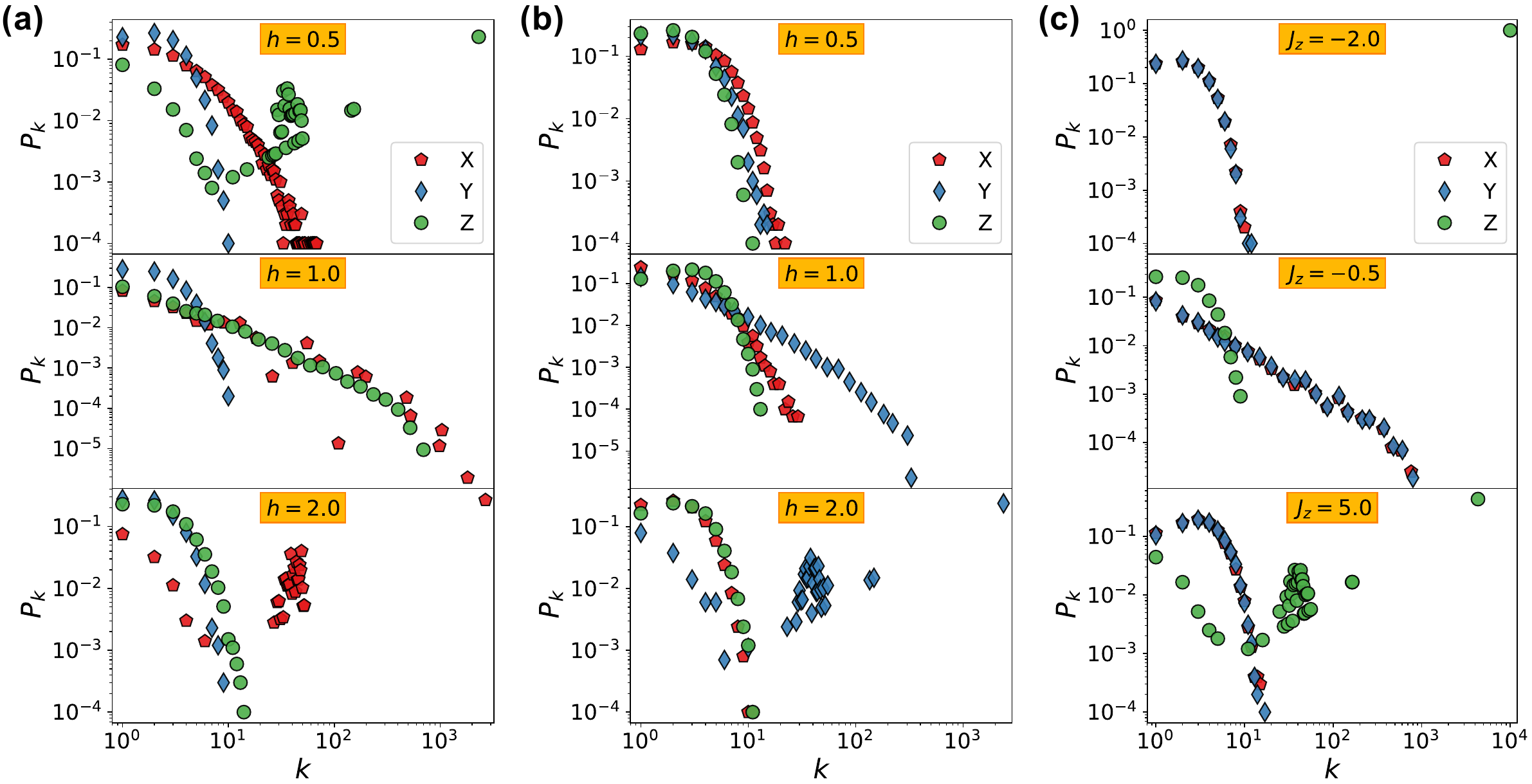}
    \caption{{\bf Degree distributions of wave function networks built from simulated ground states.} Each network has $10^4$ nodes, and links are built, according to the prescription of Sec.~\ref{sec:WFN} by putting a cutoff at the average $3^\text{rd}$ nearest-neighbor distance. Configurations are sampled in $x$-, $y$-, and $z$-basis from the ground states of the three models under examination, with $L=80$: \textbf{(a)} Quantum Ising model, \textbf{(b)} cluster Ising model, and \textbf{(c)} XXZ model. In general, we observe short-tailed distributions, typical of random networks. 
    A first exception are the FM, AFM, and PM not-critical phases in minimal complexity basis, exhibiting probability networks. These are networks consisting of isolated islands, such that their degree distributions are nothing but histograms of  repetitions, given the cutoff $R=\overline{d_3}<1$. As a second exception, networks are scale-free in the minimal-complexity basis at critical points and in critical phases. The degree distributions are linear-binned in all cases except when considering the minimal-complexity basis in critical points and phases, where they are log-binned.}
\label{fig:Results_Net}
\end{figure*}

In this section we show the network analysis we performed on the datasets sampled from the ground states we simulated. Following the procedure explained in Sec.~\ref{sec:WFN}, the networks are built out of datasets made of $N_s=10^4$ configurations each, and the resulting degree distributions are analyzed. We note that here we do not break the degeneracy of repeated  configurations, as done in the previous section, since the present analysis is not affected by such repetitions. 

In Fig.~\ref{fig:Results_Net}(a), we show the degree distributions of the networks coming from quantum Ising ground states. For $h=0.5$, in the ferromagnetic phase, we observe two distinct behaviors for the minimal-complexity basis ($z$) and for the other two bases. In the latter case, we observe short-tailed degree distributions, characteristic of random-like networks.
In the former, however, the produced network is a probability network, as introduced in Sec.~\ref{sec:Construction_WFN}.
Analyzing this histogram from a physical perspective, we find it to be consistent with our expectations: the fully magnetized configuration has the highest degree, and configurations with an increasing number of excitations have lower and lower degrees. An analogous situation is observed for $h=2.0$, in the paramagnetic phase, with $x$ being the minimal-complexity basis. Notably, at the critical point $h=1.0$, we observe that the degree distributions in the minimal-complexity bases, $x$ and $z$, are fat-tailed and compatible with a power law, a hallmark feature of scale-free networks. Note that for such cases, the degree distribution presents a bending  at a high degree $k$. This aspect is due to the finite size of the networks, as analyzed in more detail below. The network in the non-minimal-complexity basis is a random one. This is consistent with our predictions exposed in Sec.~\ref{sec:Classification}.

The networks built with the ground states of cluster Ising and XXZ models are portrayed, respectively, in Figs.~\ref{fig:Results_Net}(b) and~\ref{fig:Results_Net}(c). The observations are once again aligned with the predictions of Sec.~\ref{sec:Classification}. Specifically, the SPT phase of cluster Ising displays random networks in all bases. Then, the AFM and FM phases of both models show repetition histograms in the minimal-complexity basis and random networks in the other bases. Moreover, once again we observe signs of scale-freeness both at the critical point of cluster Ising and within the PM critical phase of XXZ, for which we present in the figure a single representative point at $h = -0.5$.

\subsection{Finite-size scaling test of scale-freeness in critical systems}

\begin{figure}
    \centering
    \includegraphics[width=\linewidth]{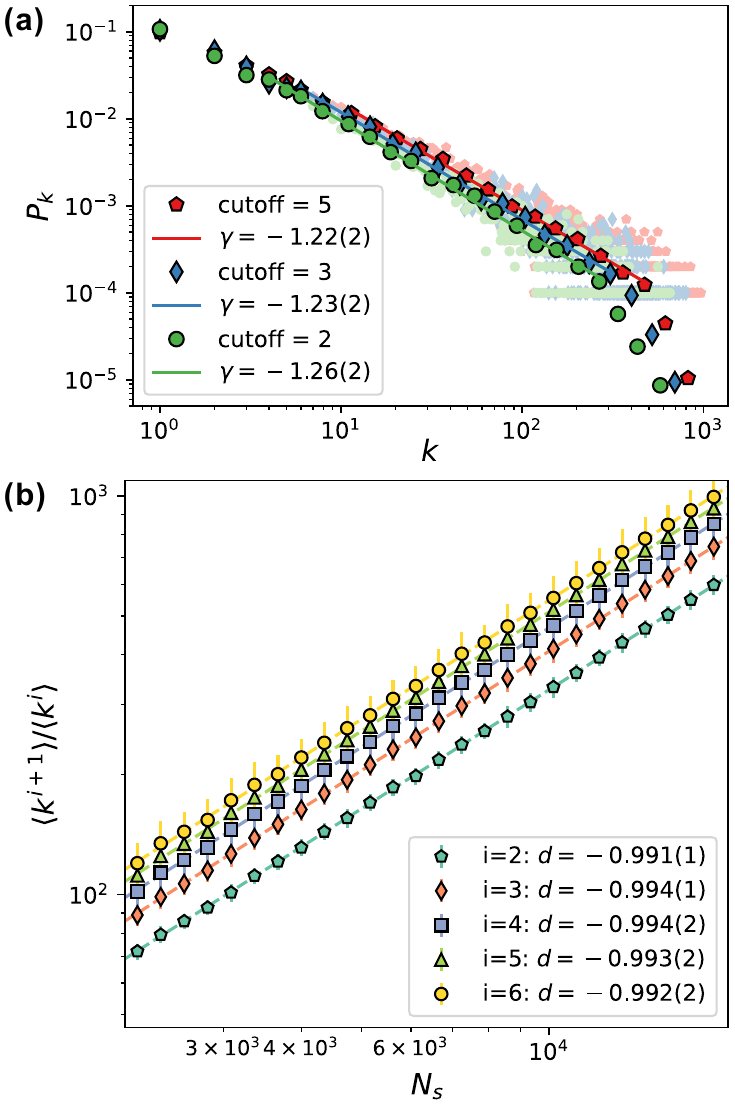}
    \caption{{\bf Assessing scale-freeness in finite-size wave function networks at quantum Ising's critical point.} \textbf{(a)} Degree distributions computed from $N_s=2\times10^4$ samples in $z$ basis. Darker and lighter colors refer to log- and linear-binned data. The considered cutoff distances are the average $2^\text{nd}$, $3^\text{rd}$ and $5^\text{th}$ neighbor's distance. Up to deviations for small and large $k$, they follow a power-law $\sim k^{-\gamma}$, with $\gamma$ independent on the cutoff within $2\sigma$. \textbf{(b)} Moments' ratios vs. subnetwork size $N_s$. They all behave as power-law with the same exponents within $2\sigma$. This shows that the bending at large $k$ in panel (a) is due to finite sampling.}
    \label{fig:Emergence_SF}
\end{figure}

As noted in the previous section, degree distributions of wave function networks in critical phases, in the minimal-complexity basis, exhibit a decay compatible with a power law with departures from that behavior at large $k$, which we attribute to their finite size.
This behavior can be observed in Fig.~\ref{fig:Results_Net} as well as in  Fig.~\ref{fig:Emergence_SF}(a), where we plot again 
the degree distribution for the Ising model at the critical point ($h=1.0$), in the $z$-basis, for varying cutoff distance. 
In the latter case, we can clearly observe a region where $P(k)\sim k^{-\gamma}$. Importantly, we  note that, within that region, the exponent $\gamma$ of the fitted power law is left unchanged (within error bars) for the different cutoffs considered. However, beyond certain $k^\star$, $P_k$ does not obey such a power law decay anymore. At an intuitive level, one can understand this eventual deviation from a power law, by noting that the limited number of nodes in a network naturally introduces a cutoff scale for the maximum possible degree that can be attained. 

In order to verify this hypothesis in a rigorous way, we performed a finite-size scaling analysis, following Ref.~\cite{Serafino2021}.  
In a given network, smaller-scale representations can be obtained by removing nodes along with their associated links, where the scale is defined by the number of remaining nodes, $N_s$. In finite-size scale-free networks, the bending behavior at high degrees follows a finite-size scaling (FSS) hypothesis relative to the network scale. Specifically, the cumulative degree distribution function is given by:
\begin{equation}
    P(k,N_s) = k^{-\gamma+1} f(kN_s^d),
\end{equation}
where $d<0$. 
A straightforward approach to testing this FSS hypothesis is to analyze how the ratios of moments of $k$ depend on the network scale. The $i$-th moment is defined as:
\begin{equation}
    \langle k^i \rangle = \int_{k_\text{min}}^\infty \dd{k} k^i k^{-\gamma} \propto N_s^{-d(i-\gamma+1)},
\end{equation}  
provided that $i>\gamma-1$. Consequently, in this regime, the ratio of consecutive moments satisfies~\cite{Serafino2021}:
\begin{equation}\label{eq:moments_ratio}
    \frac{\langle k^{i+1} \rangle}{\langle k^{i} \rangle} \propto N_s^{-d}.
\end{equation}  

Here, we test the relation in Eq.~\eqref{eq:moments_ratio} by constructing representations of the critical networks at different scales and verifying that the ratios of consecutive moments, as a function of scale, follow a power-law behavior with an exponent that remains independent of the moment order. We consider network sizes ranging from $N_s^\text{min} = 2 \times 10^3$ to $N_s^\text{max} = 2 \times 10^4$.
Since our networks are constructed geometrically, this test cannot be performed by blindly applying our standard construction method at different scales, as described in Sec.~\ref{sec:Construction_WFN}. Doing so would not yield faithful representations of a larger network at smaller scales. Specifically, as $N_s$ decreases, the computed cutoff distance would increase, introducing additional links that would not appear at larger scales.  
To address this issue, we first determine the cutoff distance for the largest network considered, $R \equiv \overline{d_3}|_{N_s^\text{max}}$. We then fix this value as the cutoff distance for networks at all scales.  

To obtain statistics, for each scale $N_s$ we produce a batch of $b=100$ networks by selecting nodes at random but without repetitions from a pool of $N_\text{tot}=10^5$ configurations. The value of the moment $\langle k^i\rangle$ at a certain scale is given by the statistical average of the moments $\langle k^i\rangle_{j=1,\dots,b}$ computed in the batch. The standard error is computed using the ``delete-$d$ Jackknife standard error estimator''~\cite{shao2012jackknife}:
\begin{equation}
    \text{SE}\simeq \sqrt{\frac{N_s}{N_s^\text{tot}-N_s}} \sqrt{\frac{1}{b}\sum_{j=1}^{b}\left(\langle k^i\rangle_j - \langle k^i\rangle \right)^2}.
\end{equation}

In Fig.~\ref{fig:Emergence_SF}, we depict the ratio of moments at the paradigmatic case represented by the quantum Ising critical point. The power-law behavior is evident in log-log scale, and the exponents returned by the fit are compatible with each other within error bars. Analogous results are obtained for all the other critical networks we considered. 
These results put on a firm ground the observation of scale-freeness in wave function networks of 1D quantum systems in critical phases. It would also be nice to complement this with a recently proposed alternative RG scheme, that has reported impressive results in the context of 2D and 3D Ising models~\cite{brunner2025snapshotrenormalizationgroupquantum}.

\section{Conclusions and outlook}

We have presented an interpretable, generic, and practical framework to analyze the stochastic sampling of many-body wave functions. Our approach relies on key capabilities of quantum simulators and computers: direct access to the many-body wave function, and control over global basis rotations. Most importantly, and contrary to generic tomographic schemes, it exploits the fundamentally limited sampling capabilities of such devices (number of samples much smaller than Hilbert space dimension), via a combination of complexity and geometrical arguments. First, by systematically detecting the basis whose sampling is described by the least amount of variables (Kolmogorov complexity), and second, by embedding snapshots in a metric space whose distance is determined by Parisi two-replica overlap. This last step allows one to replace ensembles of snapshots with networks, thus unleashing the toolbox of network theory for a deeper and systematical evaluation of quantum correlations. 

Our framework allows to classify phases of matter in a deterministic manner, based on their sampling. In one spatial dimension, the classification of translational invariant states is complete, and encompasses both gapless and gapped phases (contrary to other schemes that typically refer to either one or the other). We have demonstrated and quantified our classification in the context of several spin-1/2 models, where stochastic sampling of matrix product states is combined with both manifold and network analysis. In particular, for gapless states, a state-of-the-art scale free network analysis strongly confirms the emergence of such networks as ubiquitous in quantum matter. Taken in a single basis, our results are perfectly consistent with previous experimental observations in out-of-equilibrium systems~\cite{PhysRevX.14.021029}, as well as numerical observations in classical statistical mechanics~\cite{sun2024network}.

Based on our approach, there are multiple directions that can be explored, both at the methodological and at the conceptual level. From the methodological side, the obvious next steps are the incorporation of more advanced network science methods into the analysis of snapshots. Those include the analysis of weight-degree correlations, as well as methods that are able to diagnose topological properties - from higher-order structures, to persistent homology - that, outside the language of wave function networks, have already found interesting applications in the context of lattice field theory~\cite{PhysRevD.107.034501, PhysRevD.107.034506, PhysRevD.108.056016, 10.21468/SciPostPhys.17.4.100}. Such methods might be particularly useful not only to get finer classifications in 1D, but also to attack deconfined gauge theories in $D>1$. One promising aspect there is that the latter are related to characteristic decay of loop correlation functions (perimeter versus area law), that, for the case of topological order, are most often single-basis observables due to the abelianity of Gauss law. This may also benefit from the deep understanding of correlations in Erd\H{o}s-R\'enyi networks, that uniquely characterize topological phases. 

On the conceptual side, one immediate application of our framework would be out-of-equilibrium dynamics. Understanding the latter form an information theoretic viewpoint has been a leitmotiv of the past two decades: in particular, the recently developing framework of deep thermalization may find connection to stochastic properties of wave function networks, as it is rooted in conditional information. It would be interesting to see if ideas that combine complexity filtering with extensive correlation analysis based on networks could be informative in this respect, e.g., in view of the very challenging analysis and computational tasks involving in distinguishing different thermalization designs. Similarly, understanding wave function networks may also help clarify the connection between very powerful variational wave functions such as neural quantum states, and matter properties, as such relations are likely strongly basis dependent and have not been clearly elucidated so far. 

On the quantum information side, an interesting application may reside at the boundary between classical shadow description, and randomized measurements. While a straightforward application of our techniques to shadow sampling is likely not informative (random unitaries will generate Erd\H{o}s-R\'enyi networks in most cases), it may be possible to define hybrid frameworks where the unitary information is combined with snapshots, e.g., in the form of higher-order networks~\cite{Bianconi_2021}. 

\paragraph*{Acknowledgements. -} We thank G. Bianconi, M. Heyl, A. Laio, T. Mendes-Santos, R.~K. Panda, I. Perez-Castillo, A. Rodriguez for inspiring discussions.
M.~D. was partly supported by the QUANTERA DYNAMITE PCI2022-132919, by the EU-Flagship programme Pasquans2, by the PNRR MUR project PE0000023-NQSTI, and the PRIN programme (project CoQuS). M.~D. and R.~V. were supported in part by the ERC Consolidator grant WaveNets (Grant agreement ID: 101087692). C.M. acknowledges support from the ERC grant LATIS and EOS project CHEQS. MPS simulations employed the ITensor library~\cite{ITensor, ITensor-r0.3}.

\appendix
\section{Perfect Sampling for Matrix Product States}\label{sec:MPS_PerfSamp}
All datasets used in this work has been obtained with tensor network simulations~\cite{VVitale_Tensor_Network_package, ITensor-r0.3, ITensor}. In this appendix we briefly review this numerical approach and the sampling algorithm used on top of it.

In a tensor network, the $L$-indexed tensor $\Psi^{i_1,\dots,i_L}$, representing the state of a system of size $L$, is factorized in different tensors. In particular, for the scope of this work, we refer to the matrix product state (MPS) architecture~\cite{Fannes:1990ur, PerezGarcia_2007}, where $T$ is factorized into $L$ tensors with at most 3 indices, multiplied in a chain-like manner:
\begin{equation}\label{eq:MPS}
    \Psi^{i_1,\dots,i_L} =
    \sum_{\{b\}} A_{b_1}^{i_1}
    A_{b_1,b_2}^{i_2}
    \cdots
    A_{b_{L-2},b_{L-1}}^{i_{L-1}}
    A_{b_{L-1}}^{i_L},
\end{equation}
where $i_1,\dots,i_L$ are physical indices, referring to the local degrees of freedom, while $b_1,\dots,b_{L-1}$ are bond indices.
This expression is exact with sufficiently high dimension of the bond indices, called \emph{bond dimension}. To understand this, let's consider a product state. In this case, Eq.~\eqref{eq:MPS} is exact for bond dimension 1. If two neighboring sites live in a singlet, a bond dimension 2 is required for the corresponding bond. In this fashion, as the amount of correlation grows, a larger bond dimension is required, and the bonds are said to encode the state's correlations. Because of area-law~\cite{zeng2019quantum}, this kind of architecture is particularly suited for one-dimensional systems.

The power of this method lies in the ability to smartly compress the state by controlling the bond dimension. Indeed, one can iteratively multiply pairs of neighboring tensors, and divide them again with a singular value decomposition (SVD). Truncating the latter to $M$ biggest singular values is equivalent to truncating the state to the biggest $M$ eigenvalues of the density matrix. The compression is thus ideal in the sense of this physical interpretation.

The Hamiltonian of a system can be expressed in a tensor notation, called matrox product operator, in a similar fashion. The corresponding ground state can be retrieved via a variational algorithm called density matrix renormalization group (DMRG)~\cite{White_1993, Schollwock_2005, schollwock2011density}, where each tensor, or pairs of neighboring tensors, are iteratively optimized by solving the eigenvalue problem with a solver, e.g., the Lanczos algorithm~\cite{Lanczos_1950}. 

Once the ground state is obtained, we are interested in sampling configurations in a given basis according to Born's rule:
\begin{equation}
    p(i_1,\dots,i_L) = 
    \left|\Psi^{i_1,\dots,i_L}\right|^2.
\end{equation}
MPS simulations allow to access a powerful sampling algorithm called perfect sampling~\cite{Stoudenmire_2010, Ferris2012}, whose principle is to express the full probability as a product of conditional probabilities:
\begin{equation}
    p(i_1,\dots,i_L) = 
    p(i_1) p(i_2|i_1)
    \cdots
    p(i_L|i_1,\dots,i_{L-1}).
\end{equation}
Starting from the first site, one has to perform three steps:
\begin{enumerate}
    \item Compute $p(i_1)$ by tracing the reduced density matrix: $$p(i_1) = \sum_{i_2,\dots,i_L} \Psi^{i_1,i_2,\dots,i_L} \left( \Psi^{i_1,i_2,\dots,i_L} \right) ^\dagger;$$
    \item Sample the first degree of freedom according to $p(1)$;
    \item Project the state according to the sampled value, by contracting first tensor with the corresponding basis vector, and normalize the state.
\end{enumerate}
These three steps must be performed iteratively for all sites, in order to obtain a full configuration. Then, starting again from the initial ground state $\Psi^{i_1,\dots,i_L}$, the full procedure can be followed again, obtaining more samples unbiasedly and with no autocorrelation time.

\section{Probability networks and loops}\label{sec:Loop_nets}

\begin{figure*}
    \centering
    \includegraphics[width=0.85\linewidth]{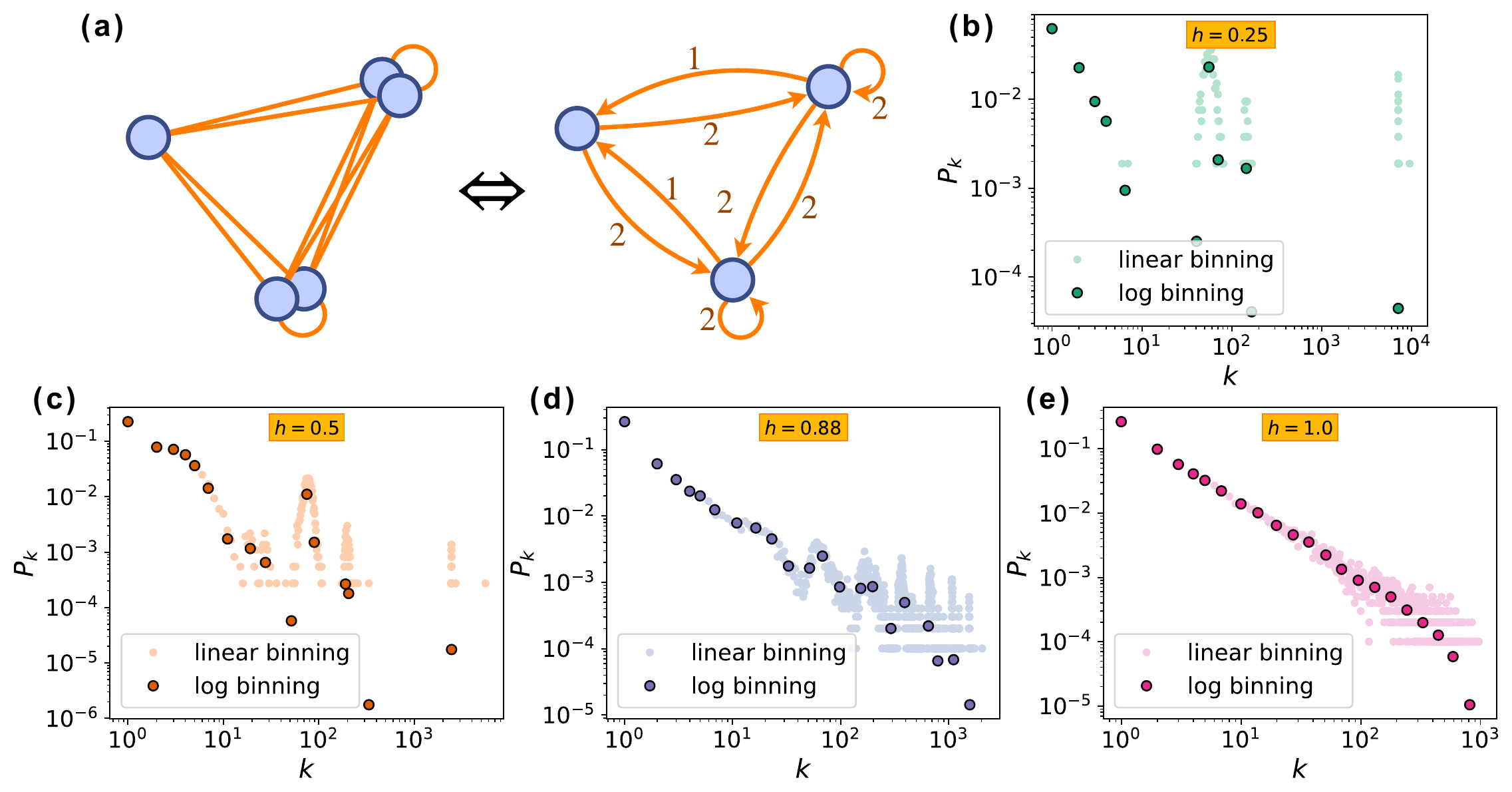}
    \caption{\textbf{Probability Networks with loops.} \textbf{(a)} Scheme of the equivalence between loop and non-loop construction. Each unique configuration is represented by a single node. Directed links are assigned using the usual geometric rule [see Sec.~\ref{sec:Construction_WFN}], and are weighted by the repetition count of the target configuration. Self-loops carry as weight their own configuration’s repetition count. \textbf{(b), (c), (d), (e)} Degree distributions of the networks for transverse field Ising model for $h=0.25, 0.5, 0.88$, in the ferromagnetic phase, and $h=1.0$, at the critical point. Each network is built with $N_s=10^4$ samples, and the cutoff distance is set as $3^\text{rd}$-nearest neighbor average distance. For the first two values, $R<1$, hence the degree distribution is the probability distribution of configurations' probabilities in the dataset. At the critical point, the dataset contains no repetitions, hence the result is the same as that of the usual procedure, in Fig.~\ref{fig:Emergence_SF}(a).}
    \label{fig:Loop_nets}
\end{figure*}

Our network construction is not the only possible one, as already stated in the text. In this section we will consider a different network construction, aimed especially at a different treatment of repeated configuration, considering networks with self loops and weighted nodes. This alternative construction is schematically shown in Fig.~\ref{fig:Loop_nets}(a).

Differently from the construction in Sec.~\ref{sec:Sec2}, here each distinct configuration in the dataset is represented by a single node, regardless of how many times it is sampled. The cutoff definition is left unchanged, and takes into account the full dataset.
The number of times a configuration has been sampled enters in the link definition: For each pair of nodes whose distance is below the cutoff distance, two directed links are considered, with a weight equal to the number of repetitions of the configuration the link is pointing towards. Moreover, For each node, a self-loop is introduced with a weight equal to the number of times the configuration itself has been sampled. Fig.~\ref{fig:Loop_nets}(b), \ref{fig:Loop_nets}(c), \ref{fig:Loop_nets}(d) and~\ref{fig:Loop_nets}(e) show the degree distributions resulting from such construction in the FM phase of transverse field Ising model.

In the limit of no repeated configurations in the dataset, the two constructions are the same. On the other hand, in the limit of a repetition-dominated sampling, where the cutoff distance becomes smaller than 1 and only self loops are present, the degree distributions are interpreted differently for this construction and the one in Sec.~\ref{sec:Sec2}. In the latter, we said that the degree distribution is an histogram of the nodes' repetitions, while in the former it represents the histogram of repetitions of the different possible configurations in configuration space. In the $N_s\to\infty$ limit, where necessarily $R\to0$, the normalized degree distribution approaches the probability distribution of configurations' probabilities for any state. 
Although the two constructions result in the same classification when applied as illustrated here, the loop construction could be suitable for different further analyses beyond this work. 

\bibliography{biblio}

\end{document}